\documentclass[twocolumn,showpacs,preprintnumbers,amsmath,amssymb,prl]{revtex4}

\usepackage{graphicx}
\usepackage{bm}

\begin{document}

\title{Water in Asbestos}

\author{Yu. D. Fomin}
\affiliation{Institute for High Pressure Physics, Russian Academy
of Sciences, Troitsk 142190, Moscow, Russia \\ Moscow Institute of
Physics and Technology, Dolgoprudny, Moscow Region 141700, Russia}

\author{E. N. Tsiok}
\affiliation{Institute for High Pressure Physics, Russian Academy
of Sciences, Troitsk 142190, Moscow, Russia}

\author{V. N. Ryzhov}
\affiliation{Institute for High Pressure Physics, Russian Academy
of Sciences, Troitsk 142190, Moscow, Russia \\ Moscow Institute of
Physics and Technology, Dolgoprudny, Moscow Region 141700, Russia}

\date{\today}

\begin{abstract}
We present the molecular simulation study of the behavior of water
and sodium chloride solution confined in lizardite asbestos
nanotube which is a typical example of hydrophilic confinement.
The local structure, orientational and dynamic properties are
studied. It is shown that the diffusion coefficient drops about
two orders of magnitude comparing to the bulk case, and water in
lizardite asbestos tubes experiences vitrification rather then
crystallization upon cooling in accordance with the results for
some other hydrophilic confinements. The behavior of sodium
chloride solutions also considered and the formation of double
layer is observed. It is shower that both sodium and chlorine have
larger diffusion coefficients then water.
\end{abstract}

 \maketitle

\section{I. Introduction}

It is well known that in spite of it's chemical simplicity water
is a very strange substance. It demonstrates a lot of anomalous
properties. That is why investigation of water is not only
important but also an intriguing field of research.

Confining water in some kind of porous materials makes the
situation even more complex. In general, confinement strongly
affects the behavior of liquids. It moves up or down melting and
boiling points, makes the density profile modulated, decreases
diffusion coefficient etc. \cite{conf1,conf2,conf3,c6h6,c6h12}
Surprisingly, in some cases the diffusion coefficient greatly
increases under confinement \cite{d1,d2,d3,d4,d5}. This phenomenon
is still to be understood.

In the case of water under confinement it is important to take
into account whether the confinement material is hydrophobic or
hydrophilic. An example of hydrophobic confinement are graphene
and carbon nanotubes. Water in this kind of confinement is widely
studied nowadays \cite{rice-review}. It is shown that in the case
of hydrophobic confinement water easily crystallizes. Sometimes it
forms the crystals of complex geometry which cannot be observed in
the bulk case (see, for example,
\cite{water-cnt,water-cnt-1,water-cnt-2}).

Even if it is of the same interest, confining water in hydrophilic
pores attracted much less attention. Mostly silica pores were
considered (see, for example,
\cite{water-sio2-1,water-sio2-2,water-sio2-3}), although some
other confining substances were studied too (for example, in Ref.
\cite{alabarse} water confined in porous $AlPO_4$ was studied).
Among the most important observations is that in hydrophilic pores
water is not inclined to crystallize. Instead of this the dynamics
of water rapidly vanishes with decrease of temperature and water
experiences a glass transition (see, for instance,
\cite{water-sio2-2,alabarse}).

One of the most hydrophilic materials is asbestos. Asbestos is a
name of a large group of natural minerals. Many forms of asbestos
form fibers. Microscopically it looks like a set of long tubes
with diameter ranging from $20$ to $50$ nm. The walls of the tubes
consist of several atomic layers and theirs width is typically
about $5$ nm. The length of the tubes can reach centimeter sizes.

In Ref. \cite{vakhrushev} an experimental study of water in
chrysotile asbestos was reported. The authors found that asbestos
quickly adsorbs drops of water from its surface. The density of
water inside the chrysotile asbestos tubes is about $1.0 g/cm^3$,
i.e. as high as bulk water. The authors found that the freezing
point of water is depressed down to approximately $237K$ and the
translation diffusion coefficient is about an order of magnitude
smaller then in the bulk case. However, they failed to measure the
rotational diffusion coefficients.

Another question of interest is the behavior of aqueous solutions
of different salts under confinement. This topic is of hot
interest for many interdisciplinary studies like, for example, the
electrolytes solutions in biological cells. Clearly, the behavior
of electrolytes at different interfaces is extremely rich and
strongly depends on the interface properties
\cite{electrolytes-interface}.

Like in case of pure water in hydrophilic confinement the aqueous
solutions were also widely studied in silica nanopores. For
example, in Ref. \cite{renou} an investigation of four different
solutions (NaCl, NaI, MgCl$_2$ and Na$_2$SO$_4$) in cylindrical
silica pore was reported. Importantly the effect of polarizability
of the force field on the behavior of the system was studied and
it was shown that in case of sodium chloride the effect of the
polarizability is negligible.

In Ref. \cite{bonnaud} a system of calcium ions in charged silica
nanopores was studied. It was found that the density distribution
of Ca$^{2+}$ ions demonstrates Stern layer but not double diffuse
layer like it should be in frames of Poisson-Boltzmann theory
\cite{electrolytes-interface}.

The influence of the ion size on the density distribution inside a
silica pore was explored in Ref. \cite{argyris}. Solutions of
sodium chloride, cesium chloride and mixture of both were studied.
It was shown that while sodium is mostly incorporated into the
second adsorbed water layer cesium predominantly stays at the pore
center. As a result cesium demonstrates much higher diffusion
coefficient than sodium even if the cesium ions are almost six
times heavier.

More publications on aqueous solutions confined into different
pores are available in the literature. To name a few, see, for
example, \cite{water-clay} for solution of sodium chloride at clay
surface and \cite{goethite} for several solutions at $(100)$
goethite surface. However, we are not aware of any study of
aqueous solutions inside asbestos tube.

From a brief overview above one can see that although the behavior
of water and aqueous solutions in asbestos fibers is of great
interest there is a lack of works in this field. The goal of the
present article is to study the microscopic properties of water
confined in lizardite asbestos tube by molecular simulation
methods. Although lizardite asbestos usually does not form fibers
it can serve as a simple model to simulate water in hydrophilic
confinement and give at least qualitative description of the
behavior of water molecules inside other types of asbestos which
form fibers, because we expect that the interaction of water
molecules with the tube walls is similar in different asbestos
types.

\section{II. System and Methods}

\begin{figure}
\includegraphics[width=8cm, height=7cm]{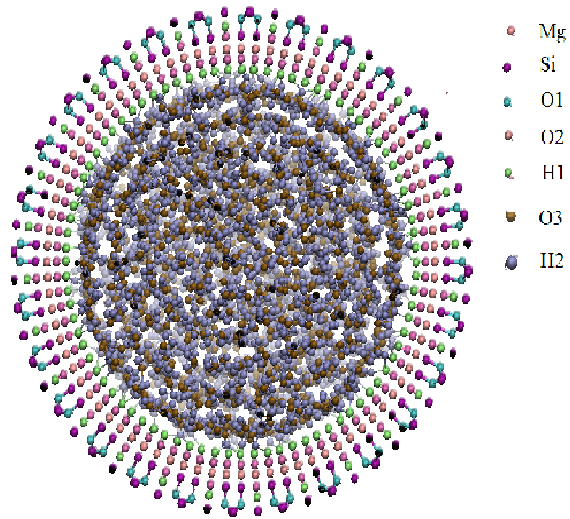}%

\caption{\label{fig:init-conf} An initial configuration for
simulation of water inside lizardite asbestos nanotube. $O1$
denotes oxygens in $SiO_4$ complexes, $O2$ - oxygens in hydroxyl
groups and $O3$ water oxygen. $H1$ is hydrogen from $OH$ groups
and $H2$ is water hydrogen.}
\end{figure}

Chemical formula of lizardite asbestos is $3MgO \cdot 2SiO_2 \cdot
2H_2O$. We consider a tube with internal radius $17.85 \AA$ and
external one $23.19 \AA$. The axis of the tube coincides with z
axis of the coordinate frame. The length of the tube is $H=53.168
\AA$. $1800$ water molecules were places inside the tube which
corresponds to the experimental density $1g/cm^3$. Initial
configuration, shown in Fig. ~\ref{fig:init-conf}, was prepared
with Packmol program \cite{packmol1}.

The tube itself is very structured. Going from the outer boundary
to the inner one we meet a layer of $SiO_4$ and $OH$ groups, the
next layer consists of magnesium and the inner one is again made
of hydroxyl groups directed with hydrogens to the inner boundary.
As a result even if the tube is electrically neutral one can see a
clear charge distribution inside the tube wall which should affect
the behavior of confined water.

We perform a set of molecular dynamics simulations of the system
at different temperatures. Clay force field was used to simulate
the asbestos atoms interactions \cite{clayff} (Mg, Si, O1 for
oxygen in $SiO_4$ groups, O2 for oxygens in $OH$ groups and H1 for
hydrogens in $OH$). SPE/E model was employed for water. Cross
interactions were obtained by Lorentz-Berthelot rule.

The system was periodic in z direction and confined in x and y
ones. In order to perform calculation of coulombic forces we
extend the simulation box in x and y directions from
$X,Y_{min}=-73.19 \AA$ up to $X,Y_{max}=73.19 \AA$ with vacuum
outside the tube and employ periodic boundary conditions in these
directions too. No corrections on the vacuum slab were made for
the sake of simplicity. We believe that these corrections are
small and do not affect the qualitative behavior of the system.

During the simulations the atoms of the tube wall were held fixed
although some simulations with free hydroxyl groups were also made
for the sake of comparison. Later on we describe it in more
details. Water molecules (and $OH$ groups when they were free to
move) were considered as rigid bodies, i.e. the bond length and
angles inside each body were fixed. The time step was set to
$0.1fs$. The system was equilibrated for $1ns$ followed by
propagation run of more $500ps$. In Refs. \cite{norman1,norman2}
general methodology of selection of time step and the standards of
molecular simulation are discussed.

We studied a wide range of temperatures starting from
$T_{min}=300K$ up to $T_{max}=1000K$ with step $dT=100K$. The
temperature was fixed by Nose-Hoover thermostat.

We also performed simulations of sodium chloride solvated in water
confined in lyzardite asbestos fibers. For doing this we took an
equilibrated structure of water inside the fiber and substituted
randomly chosen water molecules by $20$ pairs of sodium and
chlorine ions, i.e. the system contained $1760$ water molecules,
$20$ sodium cations and $20$ chlorine anions which corresponds to
concentration $0.7 mol/kg$. The simulation setup for the system
with salt was equivalent to the one of the pure water.

All simulations were performed using lammps simulation package
\cite{lammps}.

\section{III. Results and Discussion}

\subsection {Pure water}

As an initial step we consider the lyzardite asbestos tube as
being rigid, i.e. all atoms of the tube are fixed. We start the
characterization of water behavior from the radial density
profiles. The radial density is defined in the following way.
Consider a cylindrical layer with internal radius $r$ and external
one $r+dr$. The axis of the cylindrical layer coincides with the
axis of the asbestos tube. Denote as $N(r)$ the number of water
molecules inside the cylindrical layer. The position of the
molecule is taken as the position of the oxygen atom. The volume
of the layer is $V(r)= 2 \pi \left( (r+dr)^2 - r^2 \right) \cdot
H$. The radial number density is defined as $c(r)=N(r)/V(r)$ and
the radial mass density is $\rho (r) = mN(r)/V(r) $ where $m=18
a.e.$ is the mass of the water molecule. Below we use both radial
mass density expressed in $g/cm^3$ and radial number density in
$\AA^{-3}$.

\begin{figure}
\includegraphics[width=7cm, height=7cm]{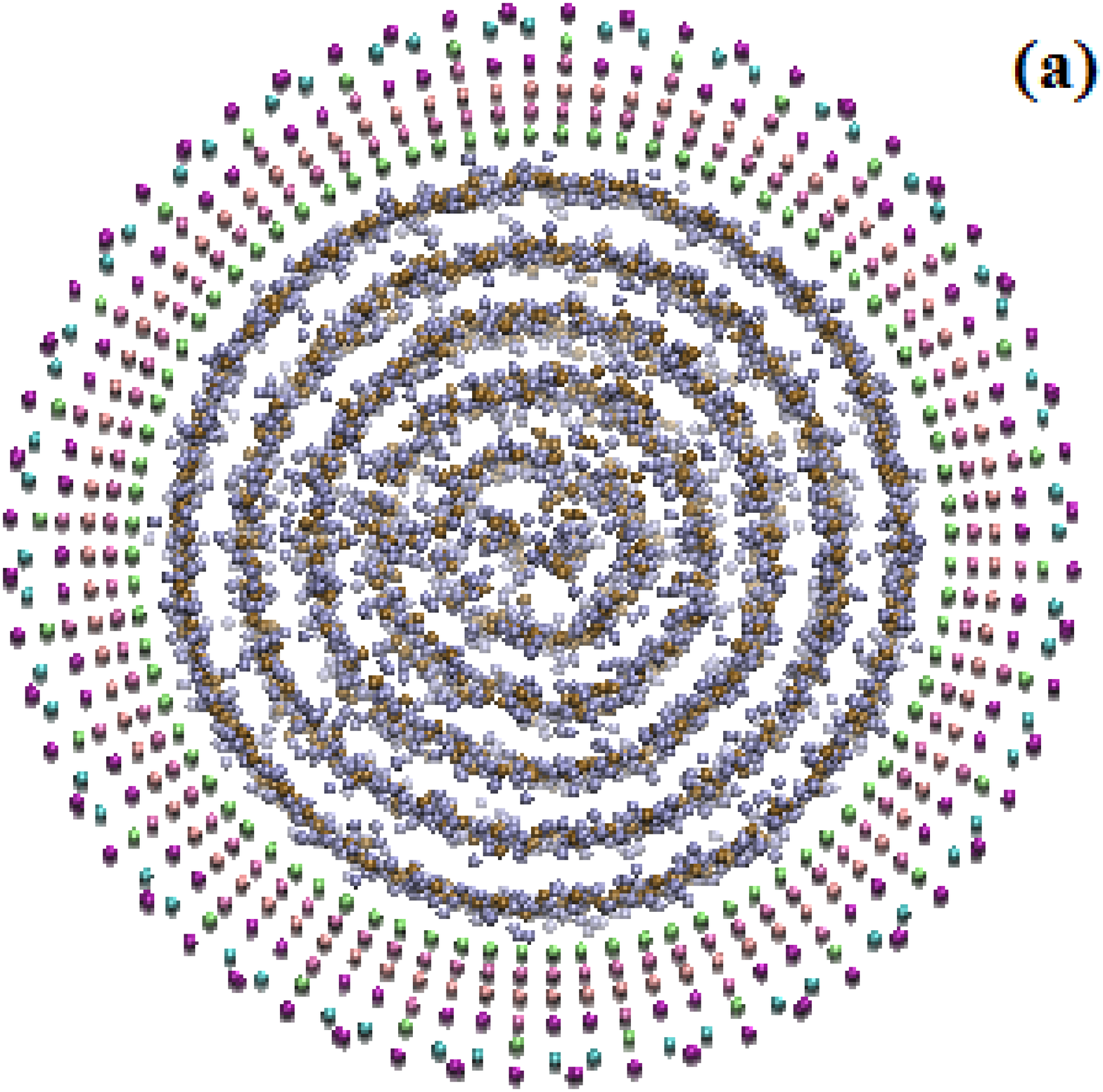}%

\includegraphics[width=7cm, height=7cm]{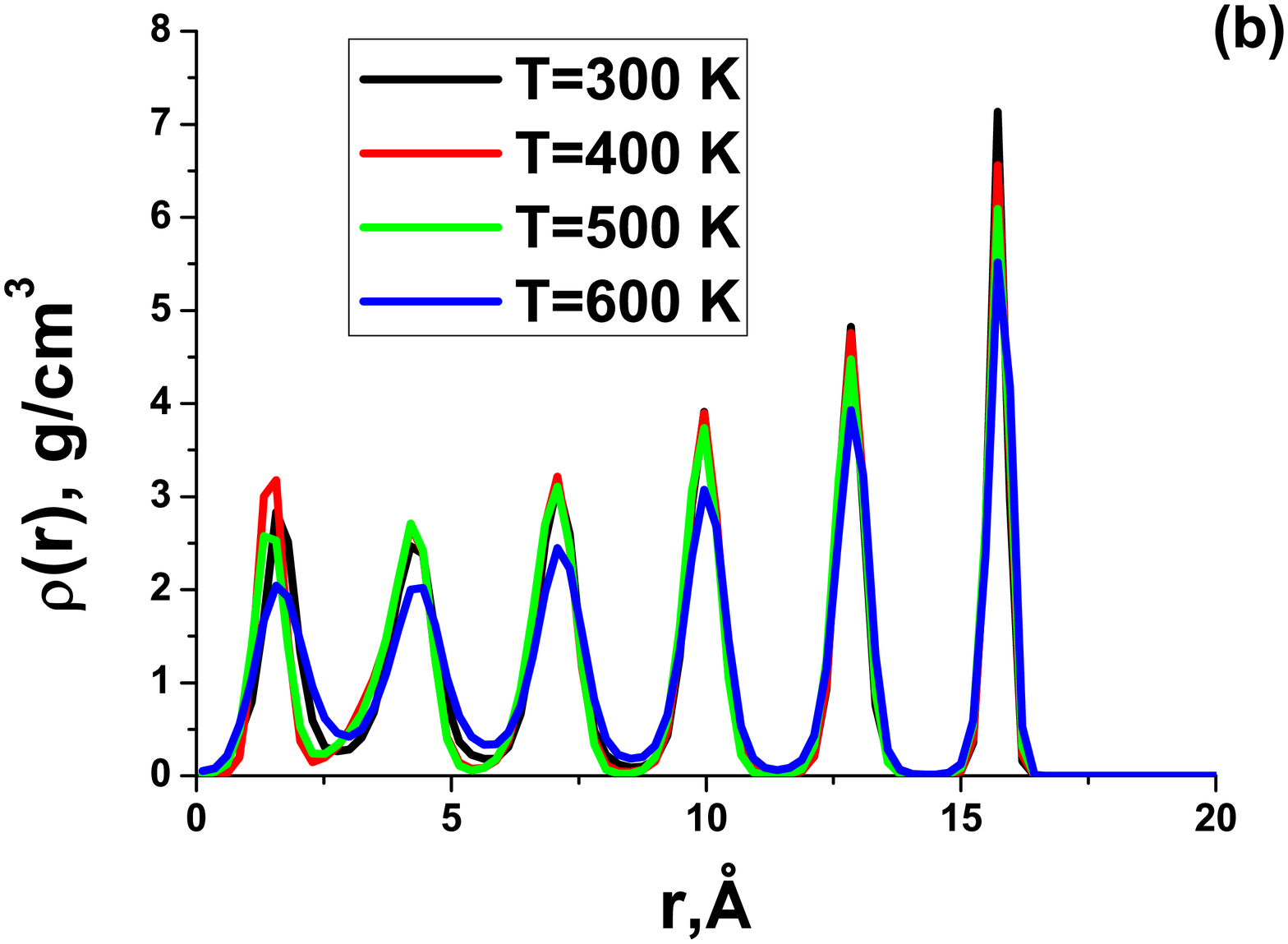}%

\caption{\label{fig:fig1} (a) Snapshot of the studied system at
$T=500K$; (b) mass density distribution profiles at temperatures
from $T=300K$ up to $T=600K$.}
\end{figure}

Fig. ~\ref{fig:fig1} (a) shows a snapshot of the system at
$T=500K$. One can see a set of very clear layers of water
molecules, i.e. the density modulations. The strength of the
modulations is as large as the radius of the tube. The internal
radius of the tube is almost $9 \AA$. Typically the density
modulations spread for $2-3$ molecular diameters, i.e. in the case
of asbestos fibers the modulations are stronger then usual. The
density profiles for the temperatures from $T=300 K$ up to $T=600
K$ are shown in Fig. ~\ref{fig:fig1} (b).

\begin{figure}
\includegraphics[width=7cm, height=7cm]{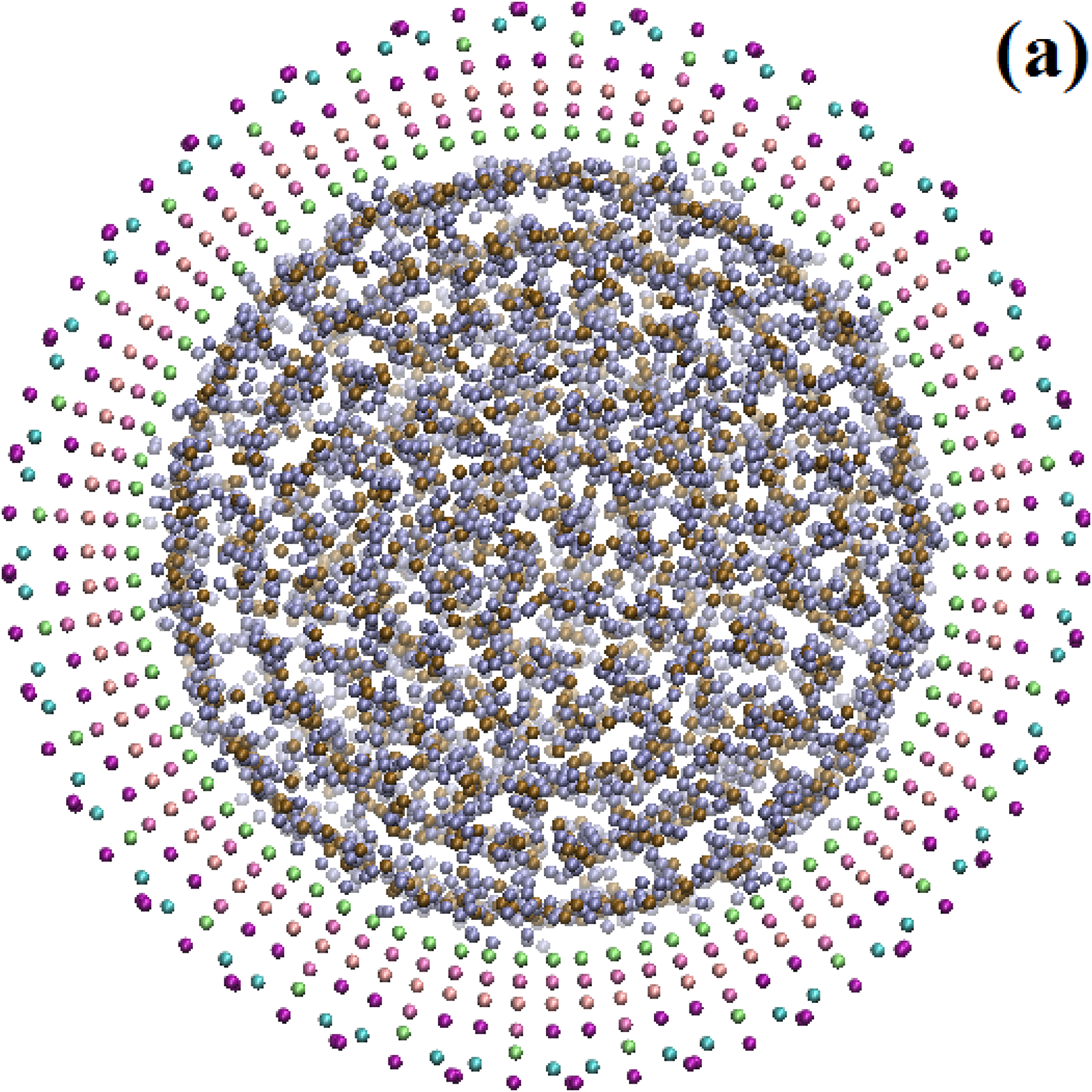}%

\includegraphics[width=7cm, height=7cm]{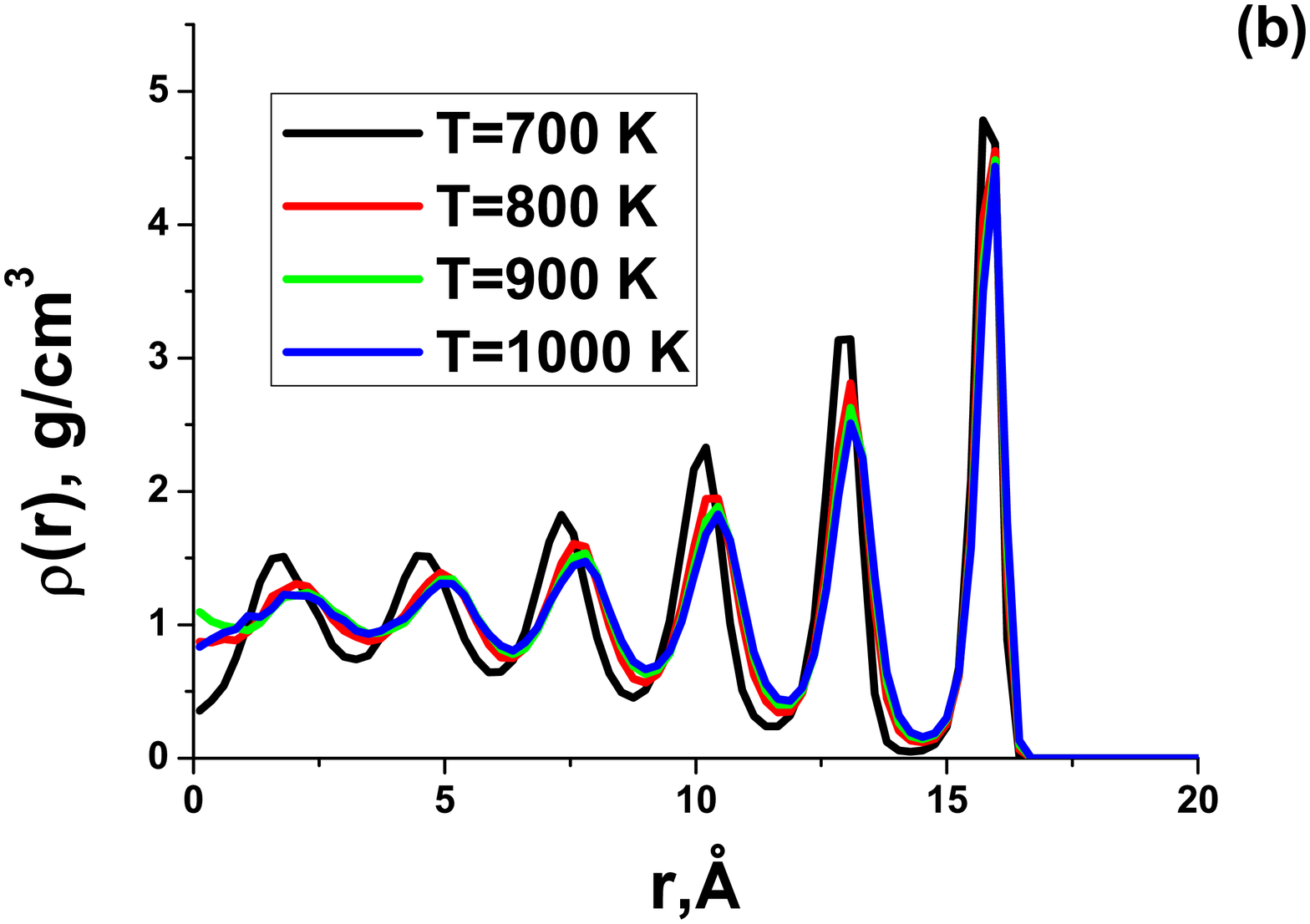}%

\caption{\label{fig:fig2} (a) Snapshot of the studied system at
$T=1000K$; (b) mass density distribution profiles at temperatures
from $T=700K$ up to $T=1000K$.}
\end{figure}

Figs. ~\ref{fig:fig2} (a) and (b) show a snapshot at $T=1000 K$
and density profiles at temperatures from $T=700 K$ up to $1000K$.
One can see that even if the structure at $T=1000K$ is strongly
smeared out it is still observable. It is even more clear from the
density profiles shown in panel (b). One can conclude that
asbestos fibers strongly modulate the density of water even at
very high temperatures.

It is also important to see wheather there is any orientational
order in different layers. Usually orientational ordering is
characterized by Legender polynomial $P_2(cos \theta)=1.5 cos^2(
\theta ) -0.5$, where $\theta$ is an angle characterizing the
molecular orientation. We chose $\theta$ as the angle between a
unit vector normal to the plane of a water molecule and the axis
of the tube. Therefore if $\theta=0$ (the plane of the molecule is
perpendicular to the axis of the tube) then $P_2 (cos \theta)=1$.
In the case of disordered orientation of the water molecules
cancellation of different contributions takes place and $P_2=0$.

\begin{figure}
\includegraphics[width=7cm, height=7cm]{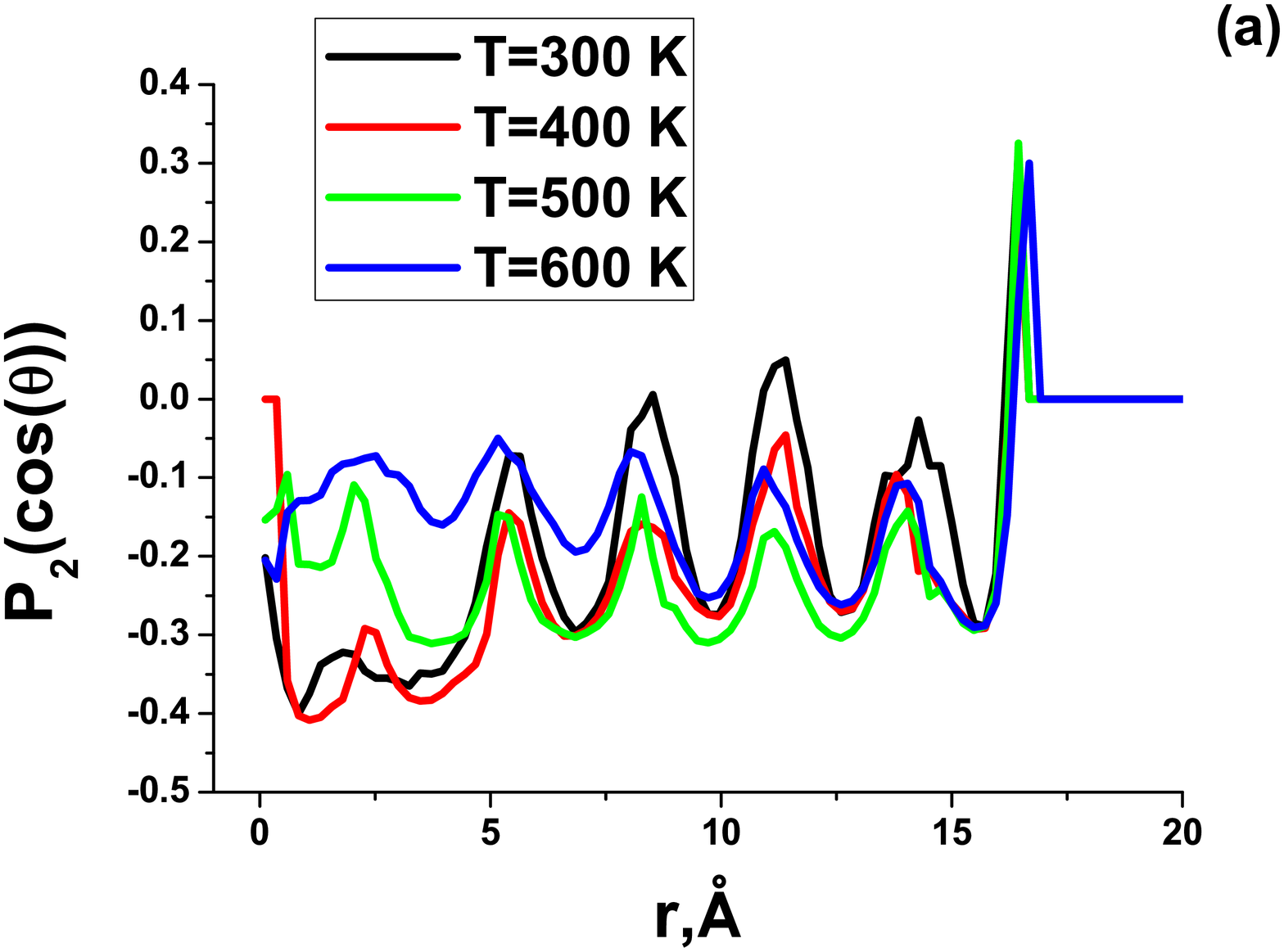}%

\includegraphics[width=7cm, height=7cm]{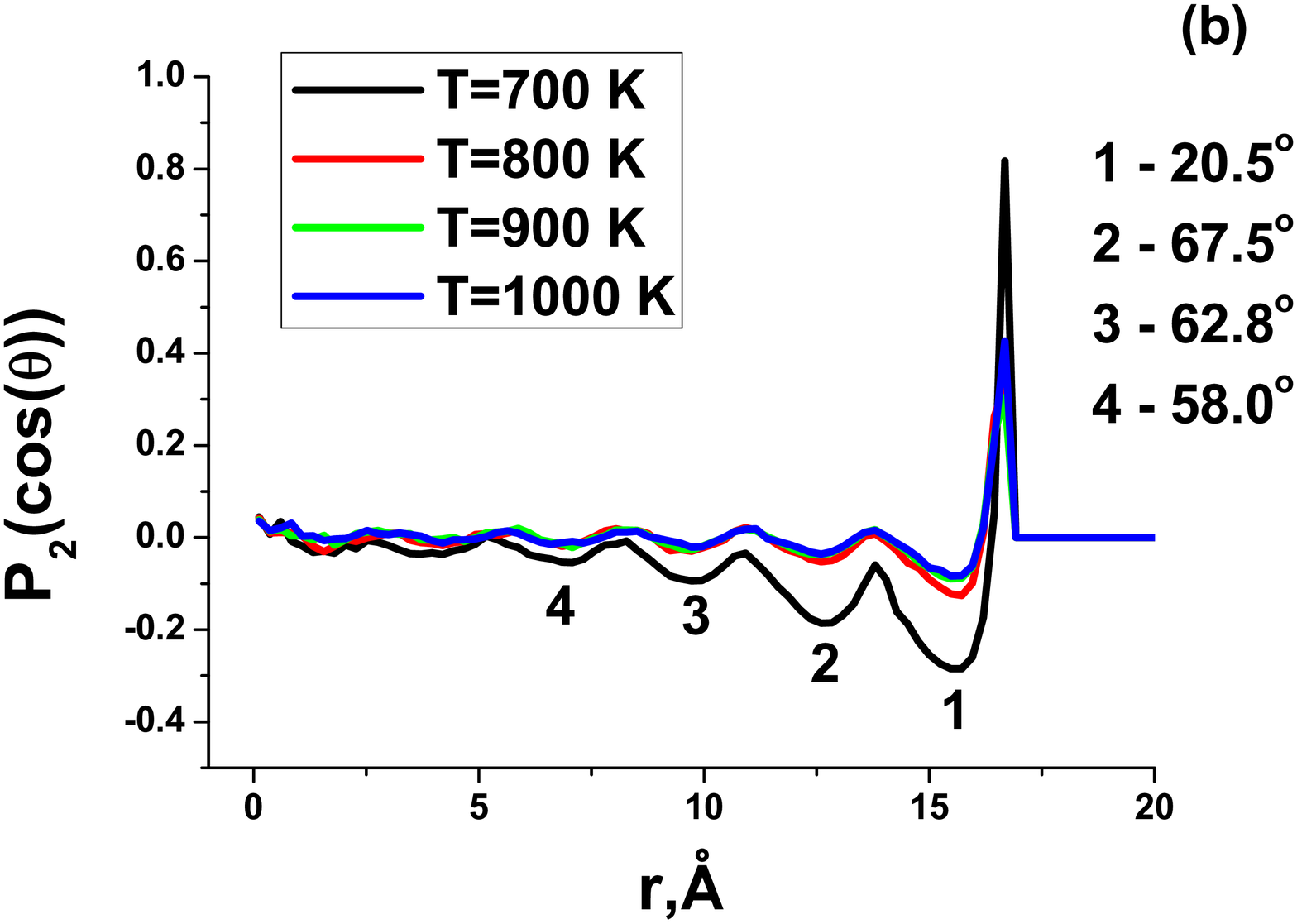}%

\caption{\label{fig:fig3} Radial distribution of Legender
polynomial $P_2(cos \theta)$ (a) for lower temperatures and (b)
for high temperatures. Numbers close to $T=700 K$ curve in panel
(b) numerate the minima at this temperature. The corresponding
angles are shown in the right side of the figure.}
\end{figure}

\begin{figure}
\includegraphics[width=5cm, height=7cm]{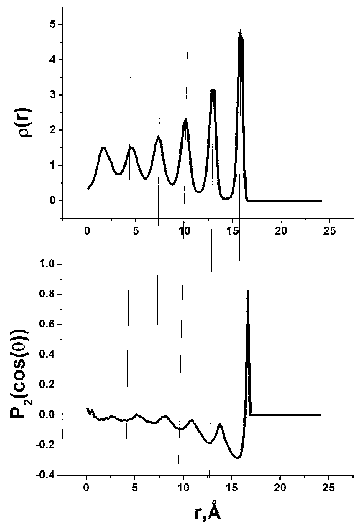}%

\caption{\label{fig:rho-p2} Comparison of the radial distribution
of the density (top panel) and $P_2(cos \theta)$ (bottom panel) at
$T=700 K$. Thin solid lines show the location of the maxima of the
density distribution.}
\end{figure}

We study the radial distribution of $P_2 (cos \theta)$ in the
system. For doing this we calculate the $P_2$ parameter for each
molecule in the layer extending from $r$ to $r+dr$ and then divide
to the number of the molecules in the layer. The results are shown
in Figs. ~\ref{fig:fig3} (a) and (b). The most pronounced
orientational structure can be observed at $T=700 K$. It
demonstrates a peak and several minima. In order to understand it
deeper let us compare the radial distribution of $P_2 (cos
\theta)$ with the radial distribution of the density at the same
temperature (Fig. ~\ref{fig:rho-p2}).

In Fig. ~\ref{fig:rho-p2} we place these distributions one below
another and draw thin lines representing the locations of the
maxima of the density distribution. One can see that zero density
corresponds to the large peak of $P_2$ close to the tube wall. It
means that the peak can be related to tiny fluctuations of the
density at this radius and most probably will be smeared out if
the statistics is increased. At the same time all minima of the
distribution of $P_2$ correspond to the maxima of the density
distribution, i.e. the higher the density the more orientated
structure is observed. It is in contrast with our recent
observations of benzene confined into a carbon nanotube
\cite{benzene-nanotube}. From the height of the main peak of
$P_2(r)$ distribution one can find that close to the wall the
molecules are oriented to $23^o$ with respect to the tube axis.

At temperatures above $700 K$ the orientational structure looks to
be smeared out due to high temperature. At lower temperatures the
structure looks odd. In order to understand it we need to consider
the dynamics of the water molecules inside the asbestos tube.

\begin{figure}
\includegraphics[width=5cm, height=7cm]{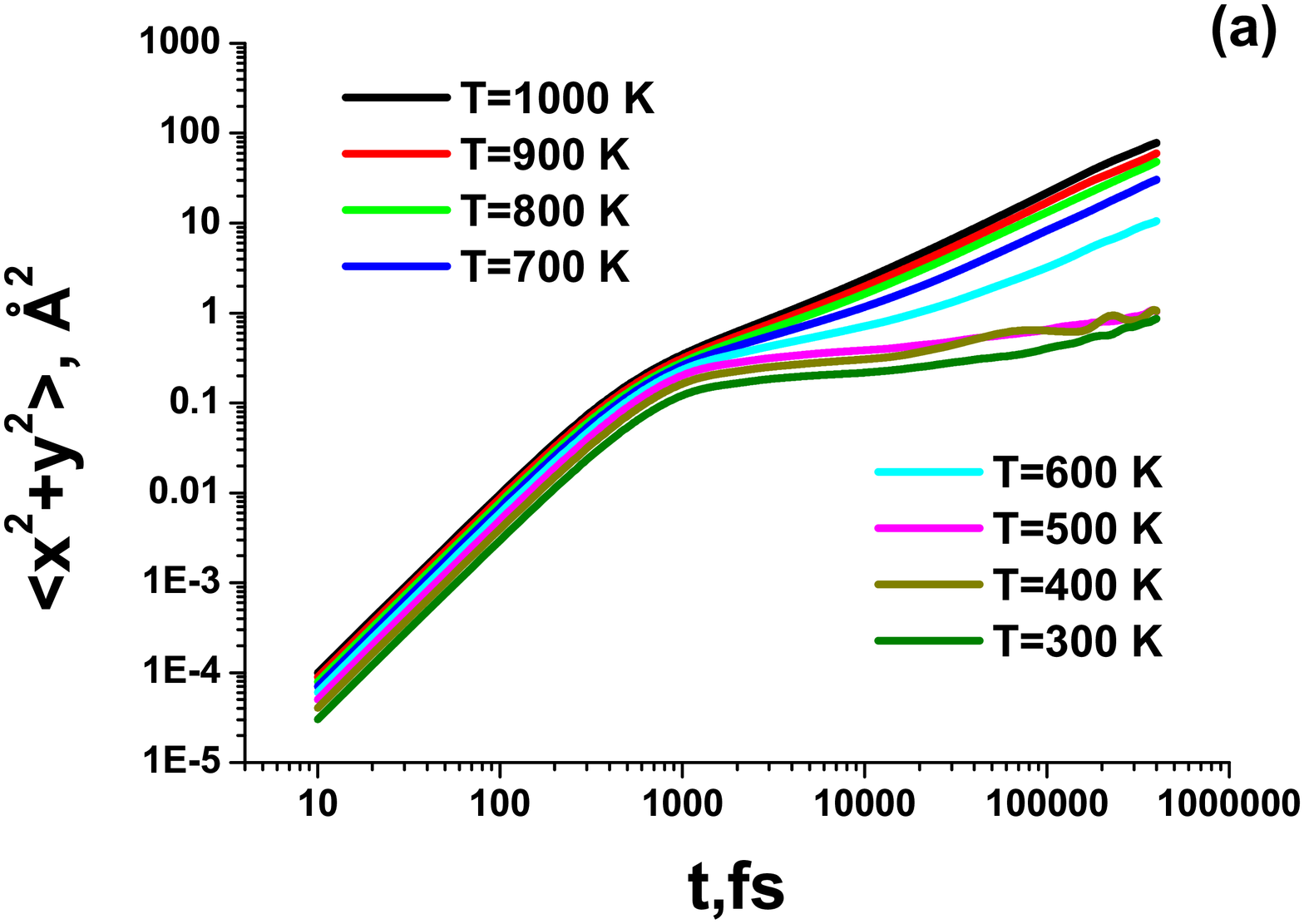}%

\includegraphics[width=5cm, height=7cm]{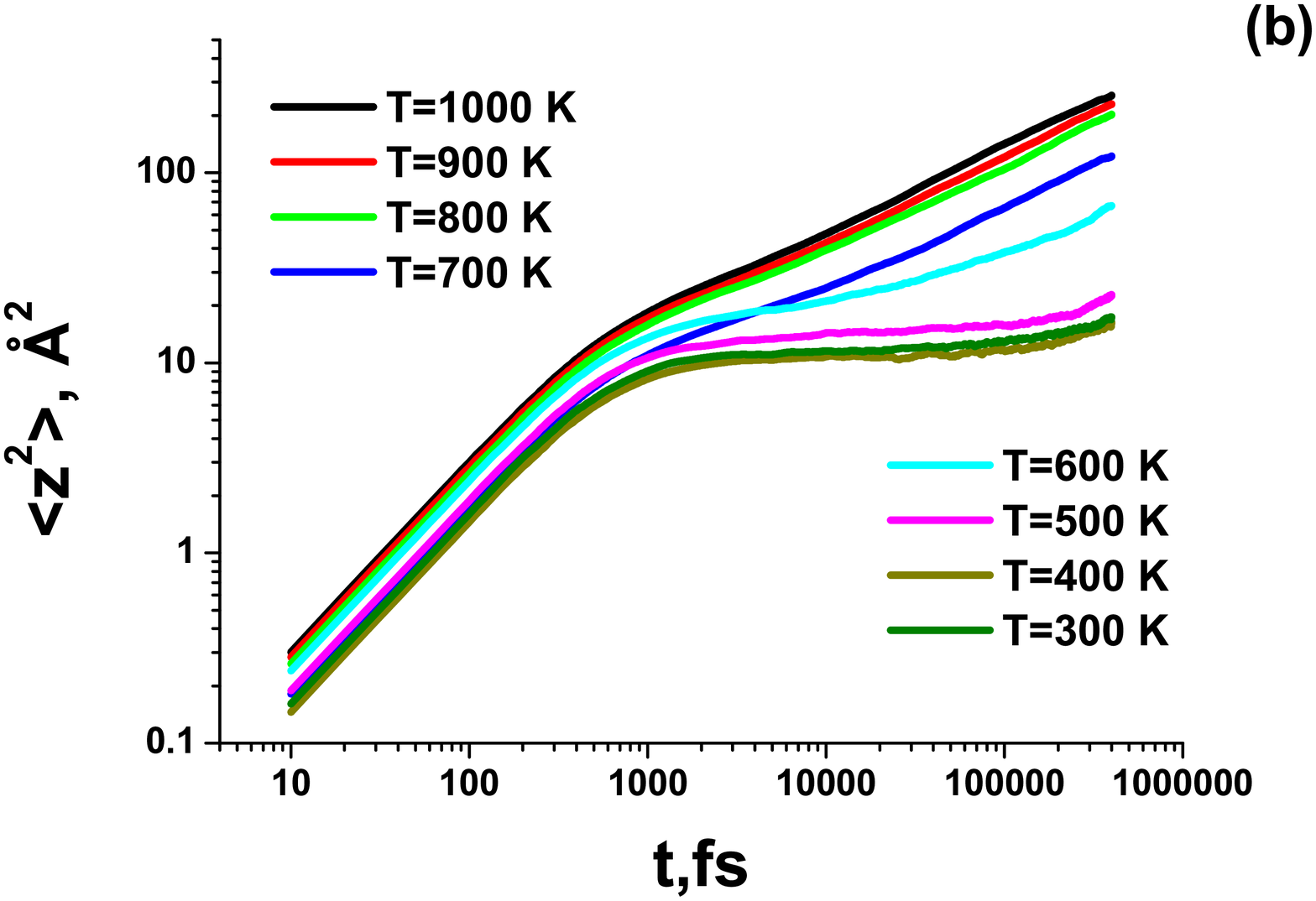}%

\caption{\label{fig:msd-water} Mean square displacement of water
molecules (a) in radial direction and (b) along the tube axis.}
\end{figure}

\begin{figure}
\includegraphics[width=5cm, height=7cm]{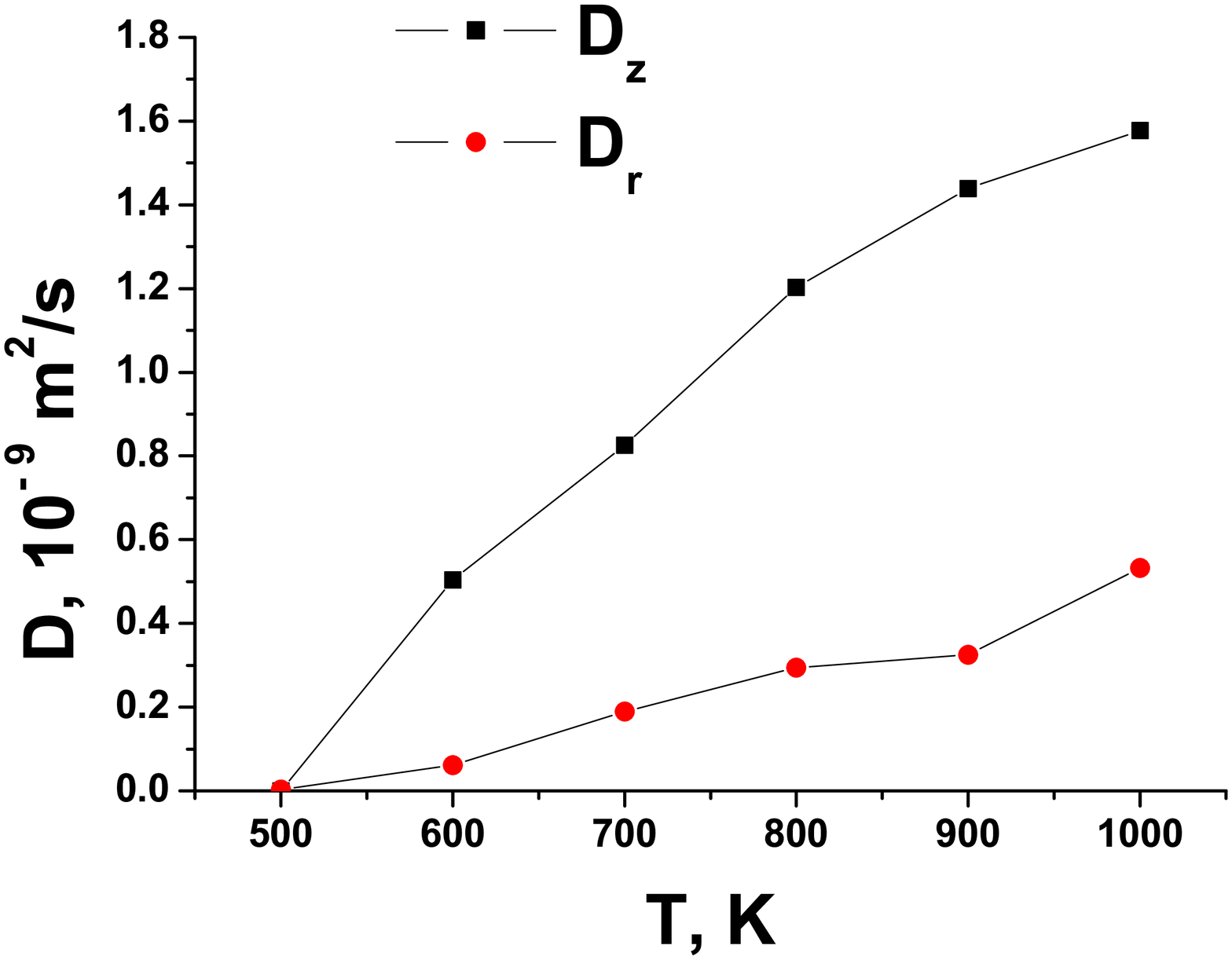}%

\caption{\label{fig:diff-water} Diffusion coefficient of water
confined in asbestos fiber. $D_r$ corresponds to the radial
diffusion coefficient and $D_z$ to the diffusion along the tube
axis.}
\end{figure}

Figs. ~\ref{fig:msd-water} (a) and (b) show mean square
displacements (MSD) in radial direction ($<x^2+y^2>$) and along
the tube axis ($<z^2>$). One can see that at temperatures about
$T=600 K$ the dynamics of the system becomes extremely slow. Fig.
~\ref{fig:diff-water} demonstrates the diffusion coefficient
computed from MSD by Einstein relation. One can see that at $T=500
K$ the diffusion coefficients become zero (for the sake of
comparison we calculated the diffusion coefficient of bulk SPC/E
water at $T=600K$: $D_{bulk}=2 \cdot 10^{-8} m^2/s$, i.e. the
diffusion coefficient of confined water is about $100$ times
smaller). Therefore the system can be considered as being trapped
in some state and cannot escape from it. It means that the results
for $T \leq 600 K$ are strongly affected by the initial
configuration which was set random and therefore the low
temperature results can be considered as some approach to the real
behavior of the system but not as exact ones.

In the present study we explore water in asbestos in the range of
temperatures ranging from $300K$ up to $1000 K$. It is known from
experimental studies (see, for example, \cite{asb-degradation})
that asbestos is thermally degradable. The degradation of asbestos
passes several stages. Firstly, water inside the fibers
evaporates. At the second step the following chemical reaction
takes place: $Mg_3Si_2O_5(OH)_4 \longrightarrow Mg_3Si_2O_7
+2H_2O$, i.e. asbestos looses the hydroxyl groups which form water
and react with the rest of the tube. The water formed from the
hydroxyl groups also evaporates. At the last step the system
decomposes to a mixture of two magnesium silicates: $Mg_2SiO_4$
and $MgSiO_3$.

If we make the $OH$ groups free we can see thermal decomposition
of asbestos at $T=500K$. We observe in our simulations that
hydroxyl groups leave theirs places in the tube and go inside it.
However, one cannot simulate chemical reactions in frames of
classical force field like the one used here. Therefore the
simulations with free $OH$ groups become unphysical at $T \geq
500K$, that is why we have to held the tube rigid.

In order to estimate the effect of the rigidity of the tube we
perform some calculations of the properties of the system at
$T=400 K$ with free hydroxyl groups.

\begin{figure}
\includegraphics[width=5cm, height=7cm]{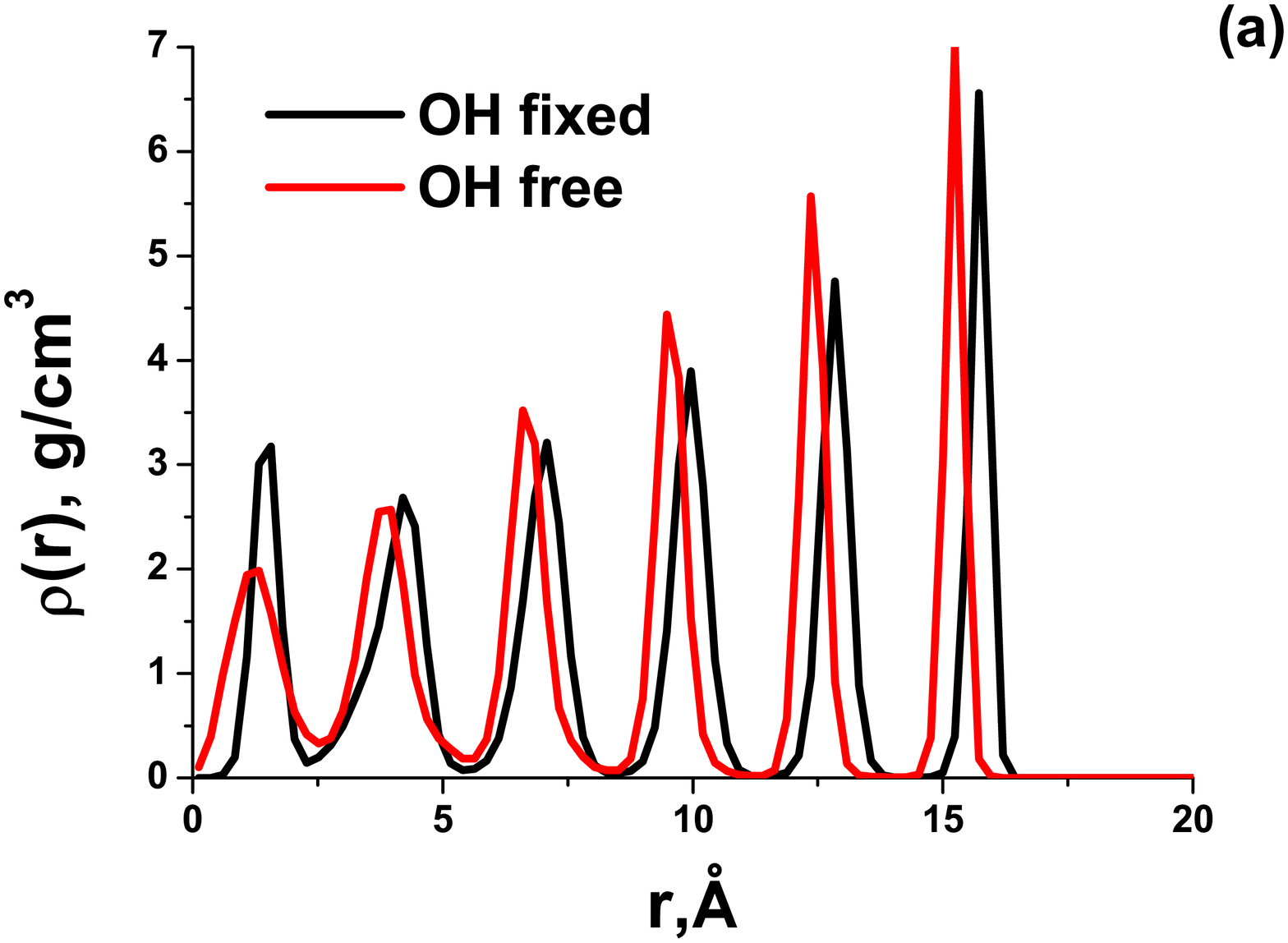}%

\includegraphics[width=5cm, height=7cm]{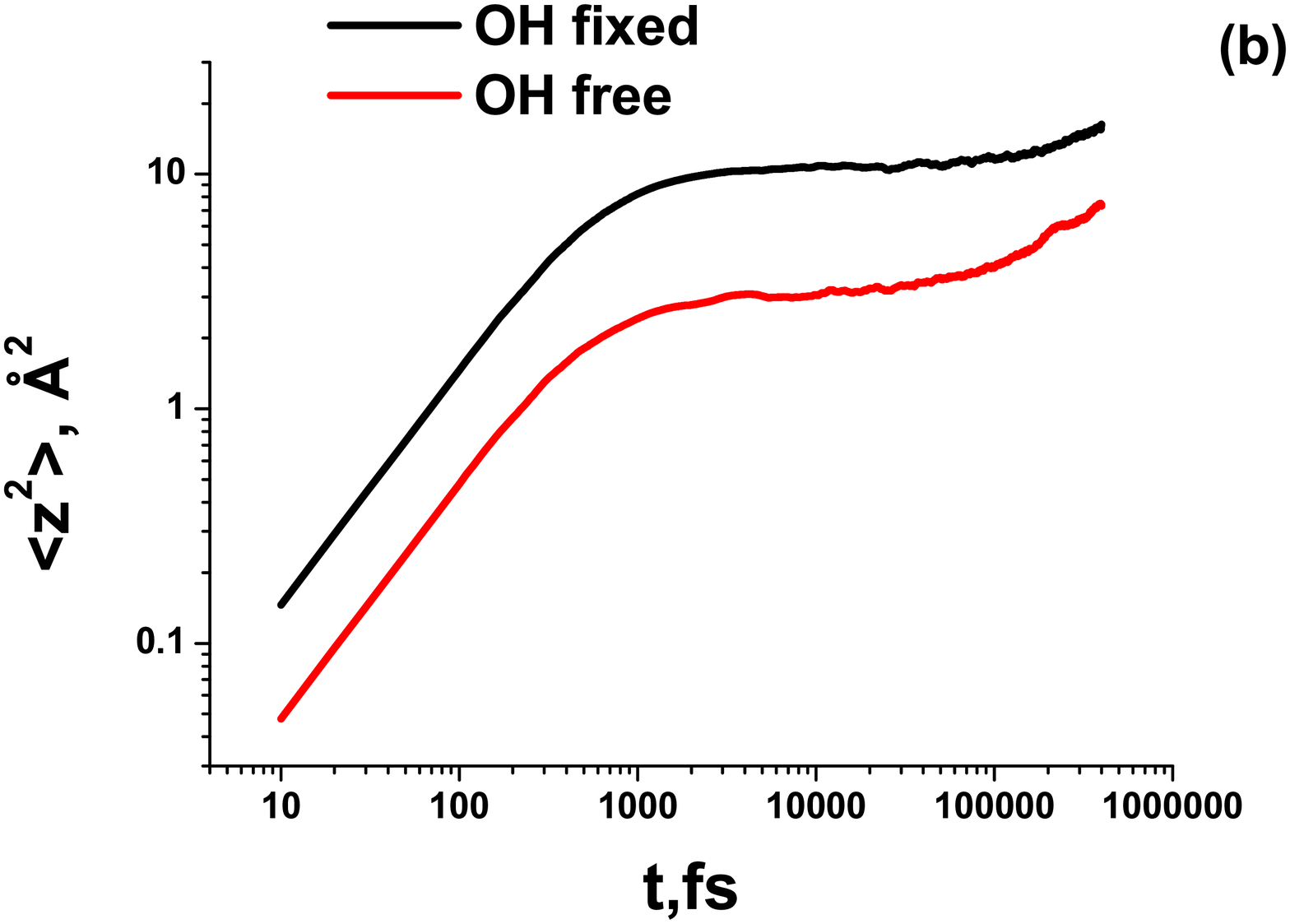}%

\caption{\label{fig:free-OH} (a) Radial distribution of density of
water inside the asbestos fiber with free and fixed $OH$ groups at
$T=400 K$; (b) mean square displacement along the tube axis for
the same conditions.}
\end{figure}

Fig. ~\ref{fig:free-OH} (a) shows radial density profiles of water
at $T=400K$ with fixed and free hydroxyl groups. One can see that
in the case of free $OH$ groups the structure becomes more
pronounced. Moreover, the positions of the peaks move toward the
central axis of the tube. However, there is no strong qualitative
differences between these cases.

Fig. ~\ref{fig:free-OH} (b) demonstrates the mean square
displacement (MSD) of water along the tube axis for fixed and free
hydroxyl groups at $T=400 K$. One can see that in the case of free
$OH$ groups MSD becomes even smaller. As it was discussed above in
the case of fixed hydroxyl groups we do not observe diffusive
behavior in the time scale of our simulations. Therefore we do not
observe it in the case of free $OH$ groups too.

\begin{figure}
\includegraphics[width=7cm, height=7cm]{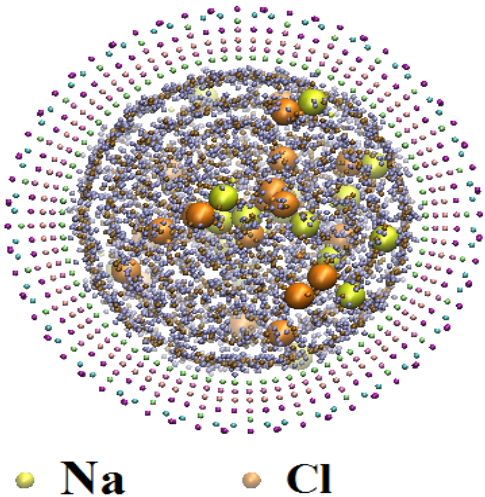}%

\caption{\label{fig:water-nacl} A snapshot of sodium chloride
solution inside the asbestos fiber at $T=900 K$.}
\end{figure}

\begin{figure}
\includegraphics[width=4cm, height=5cm]{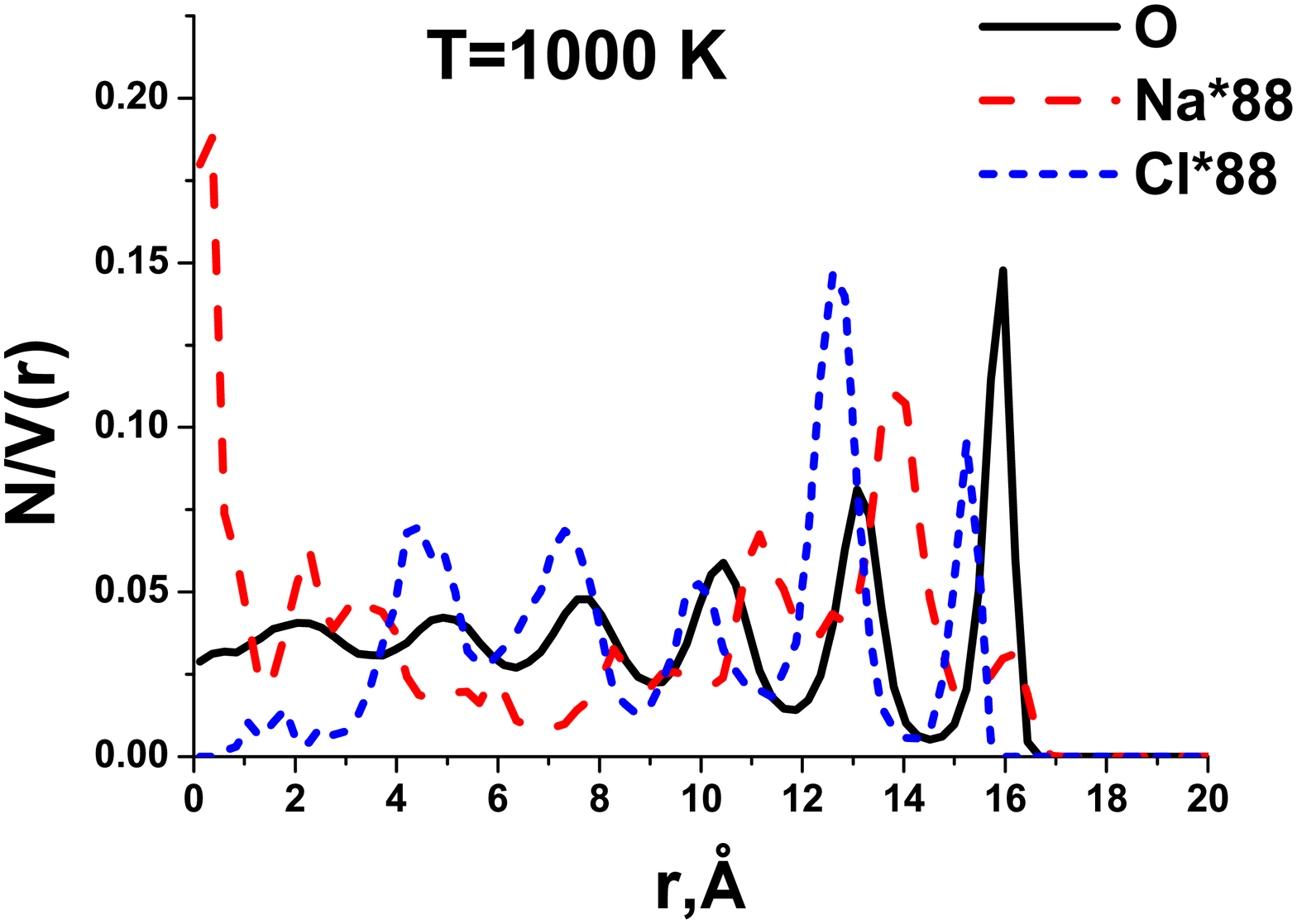}%
\includegraphics[width=4cm, height=5cm]{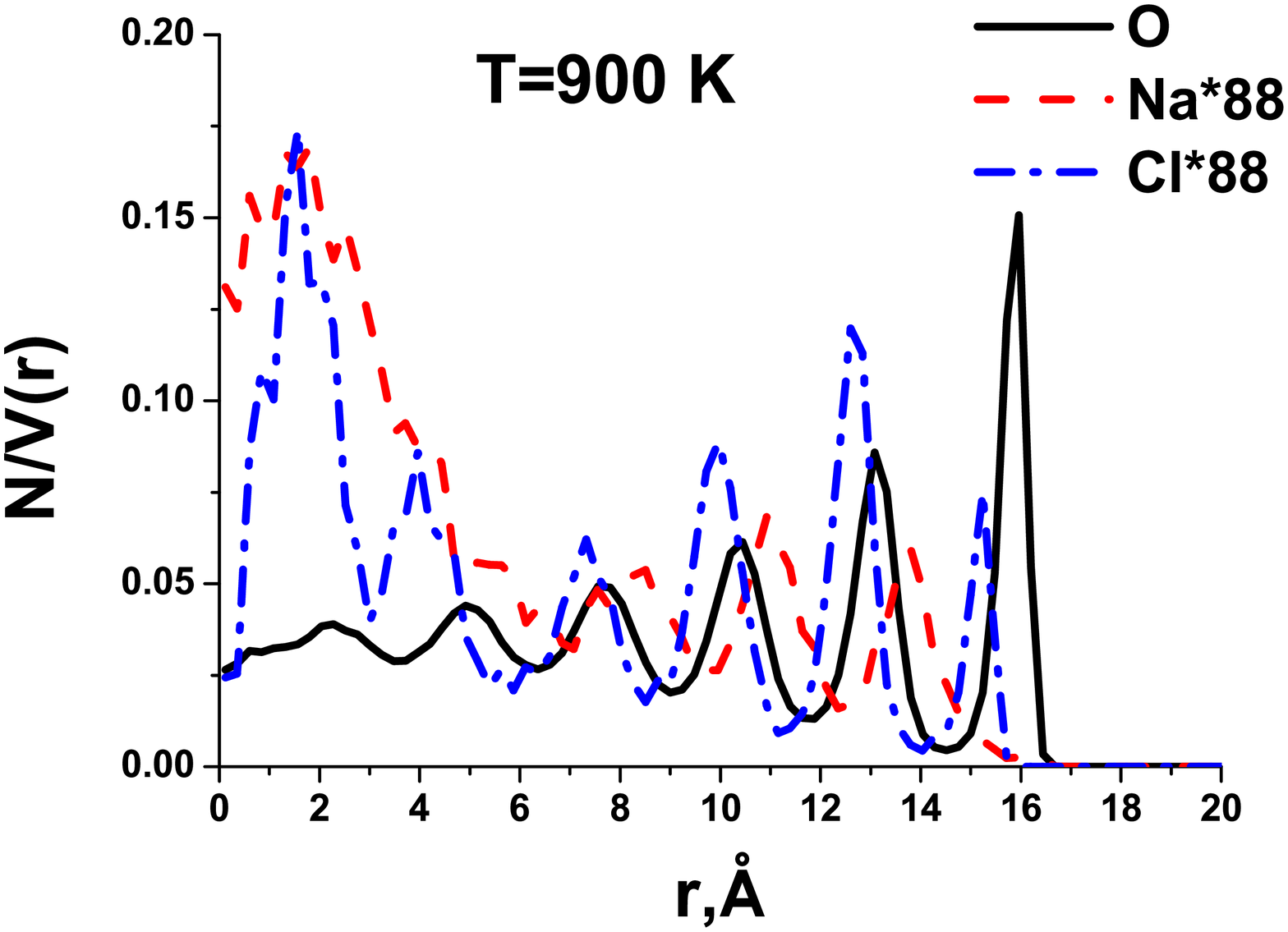}%

\includegraphics[width=4cm, height=5cm]{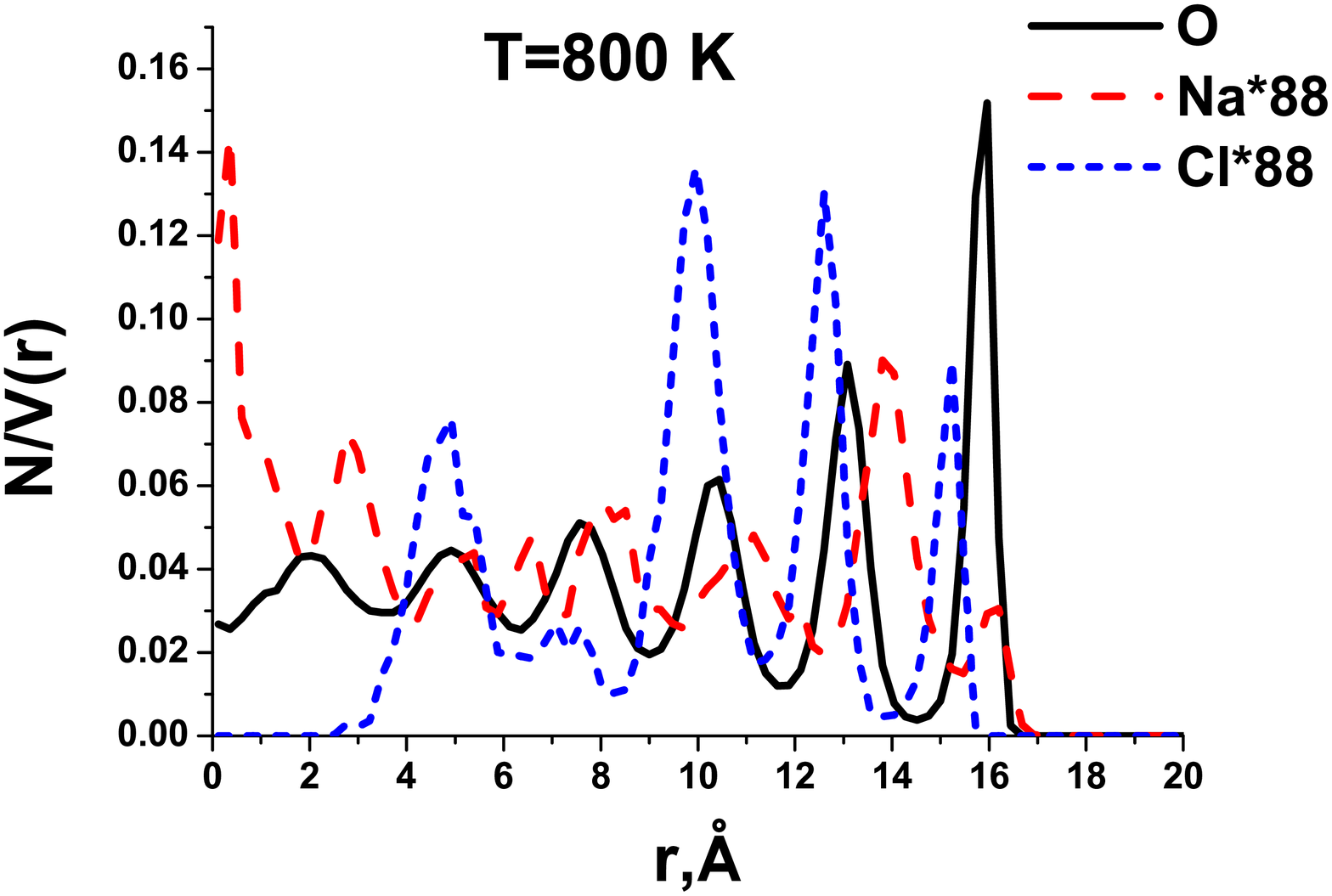}%
\includegraphics[width=4cm, height=5cm]{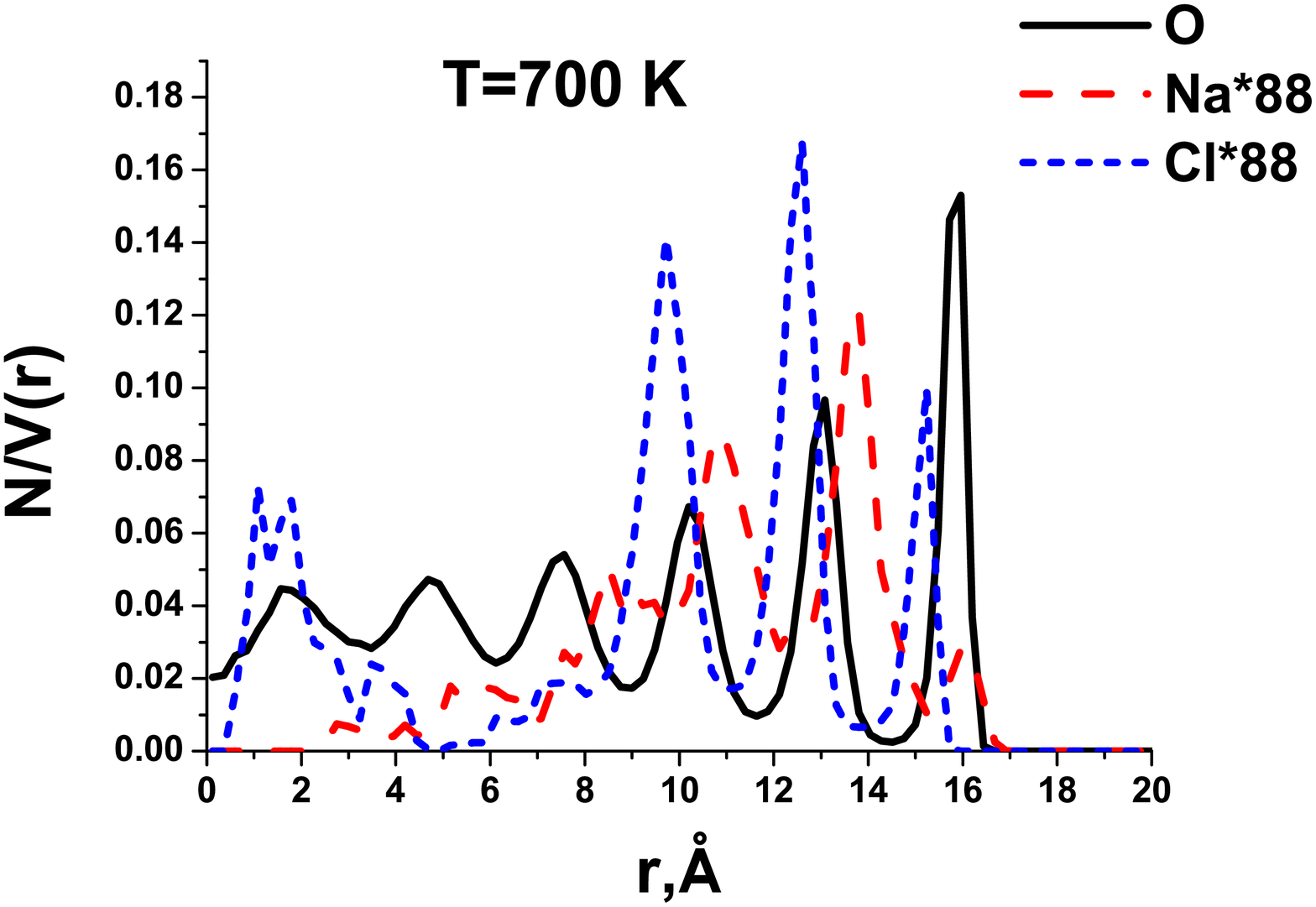}%

\includegraphics[width=4cm, height=5cm]{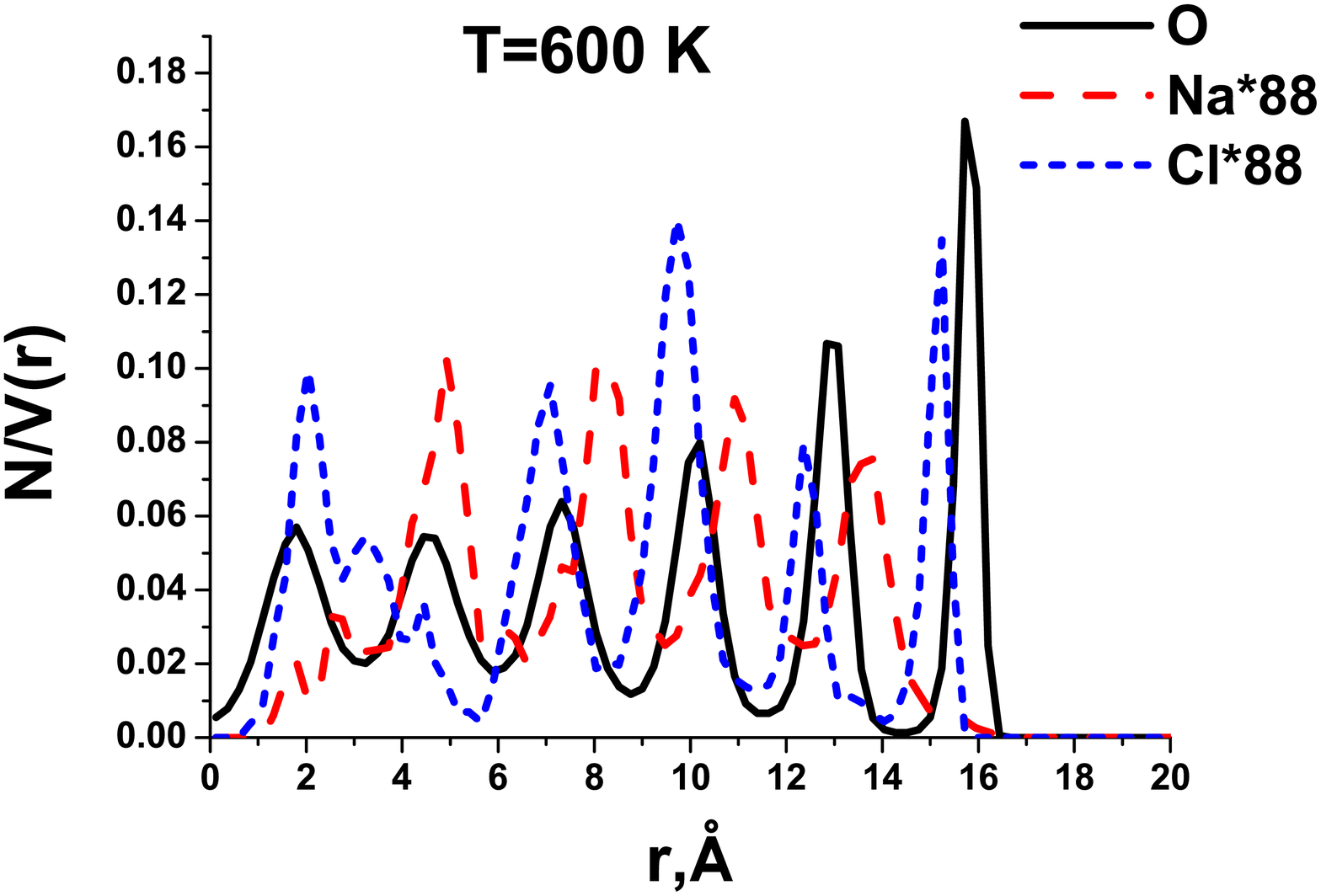}%
\includegraphics[width=4cm, height=5cm]{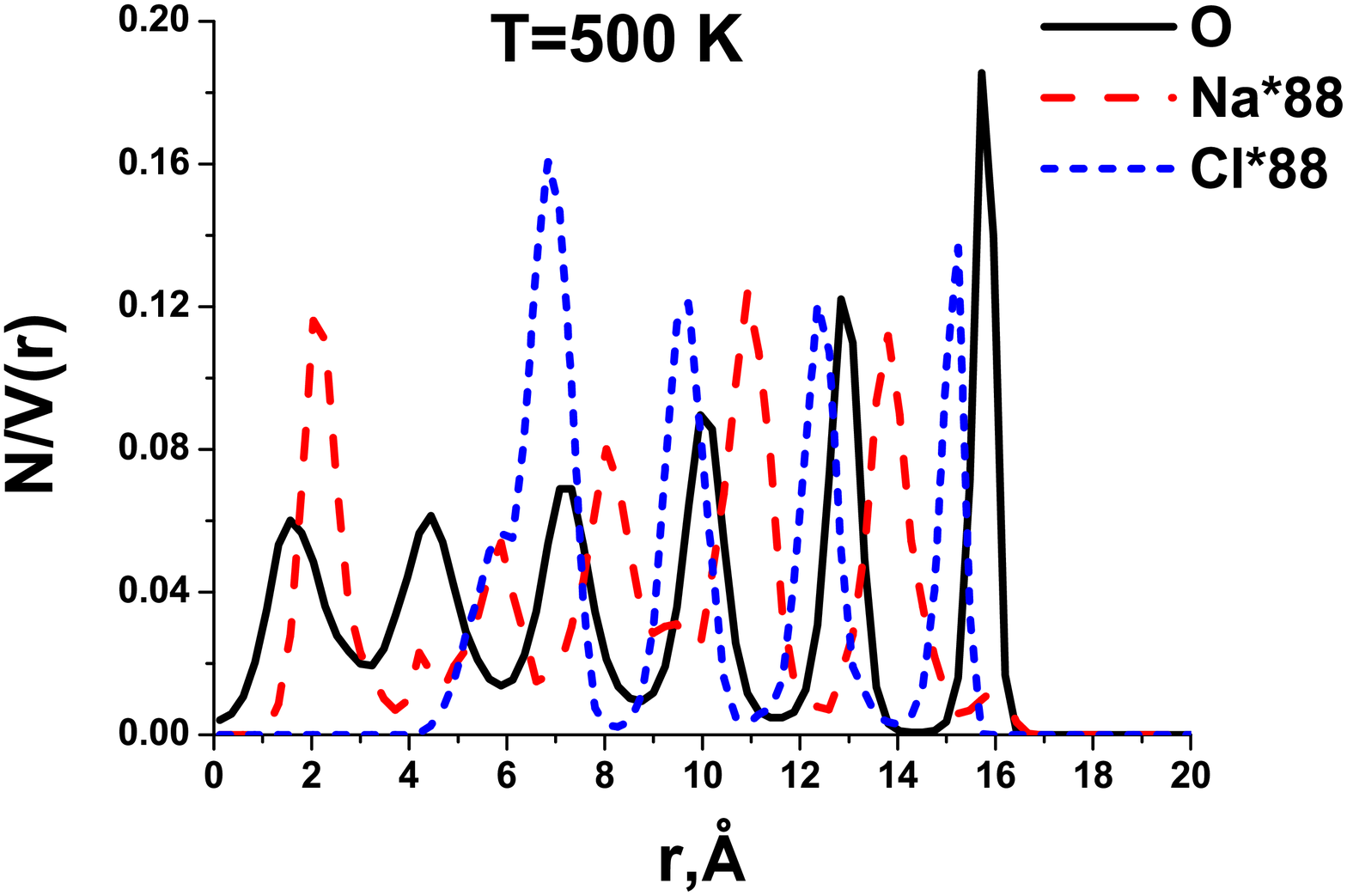}%

\caption{\label{fig:den-nacl} Radial density distribution of
different species inside the asbestos fibers. Number density $N/V$
is used. In order to have all curves in the same scale we multiply
the curves for sodium and chlorine to the factor $1760/20=88$.}
\end{figure}

\begin{figure}
\includegraphics[width=6cm, height=6cm]{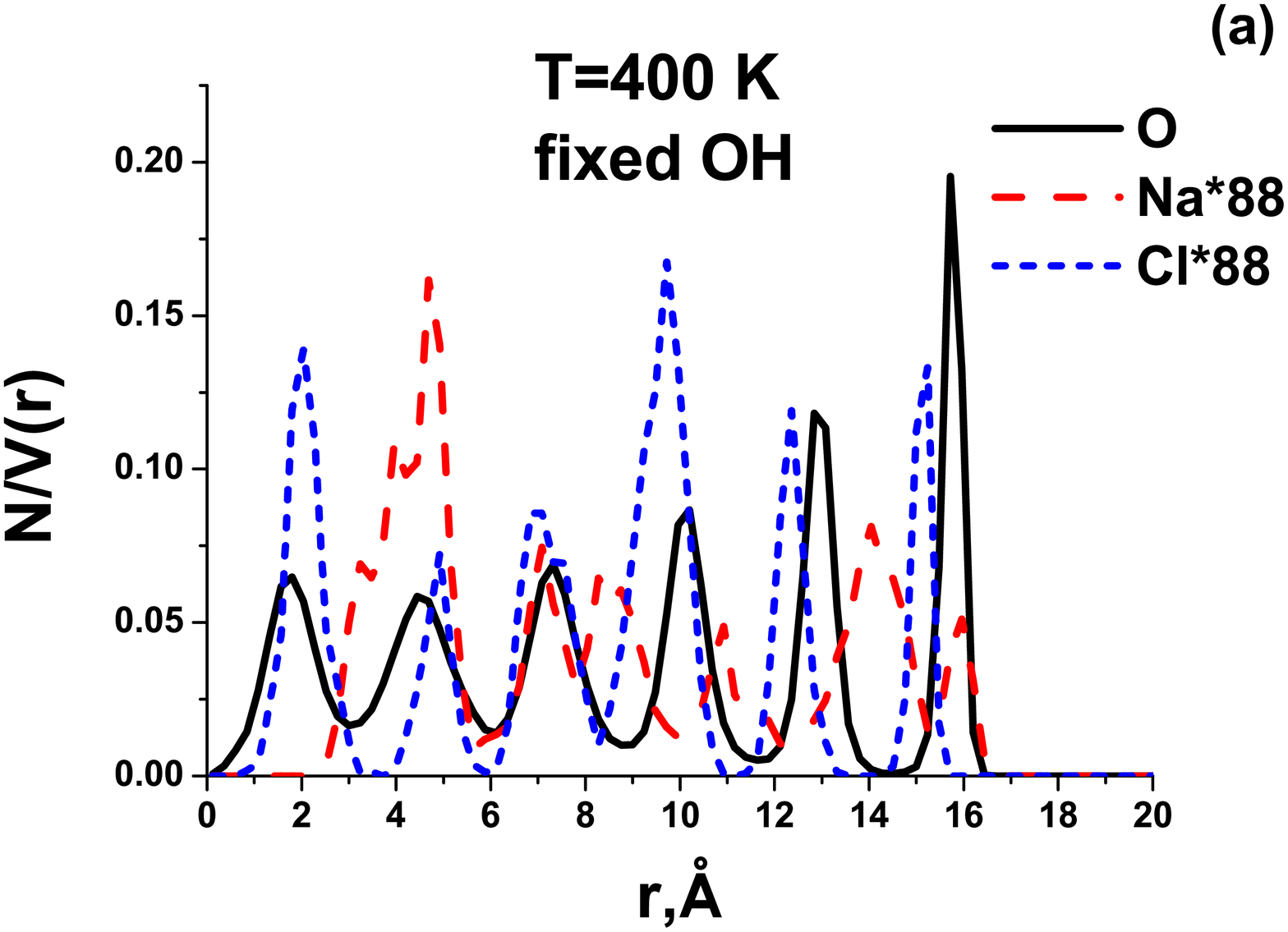}%

\includegraphics[width=6cm, height=6cm]{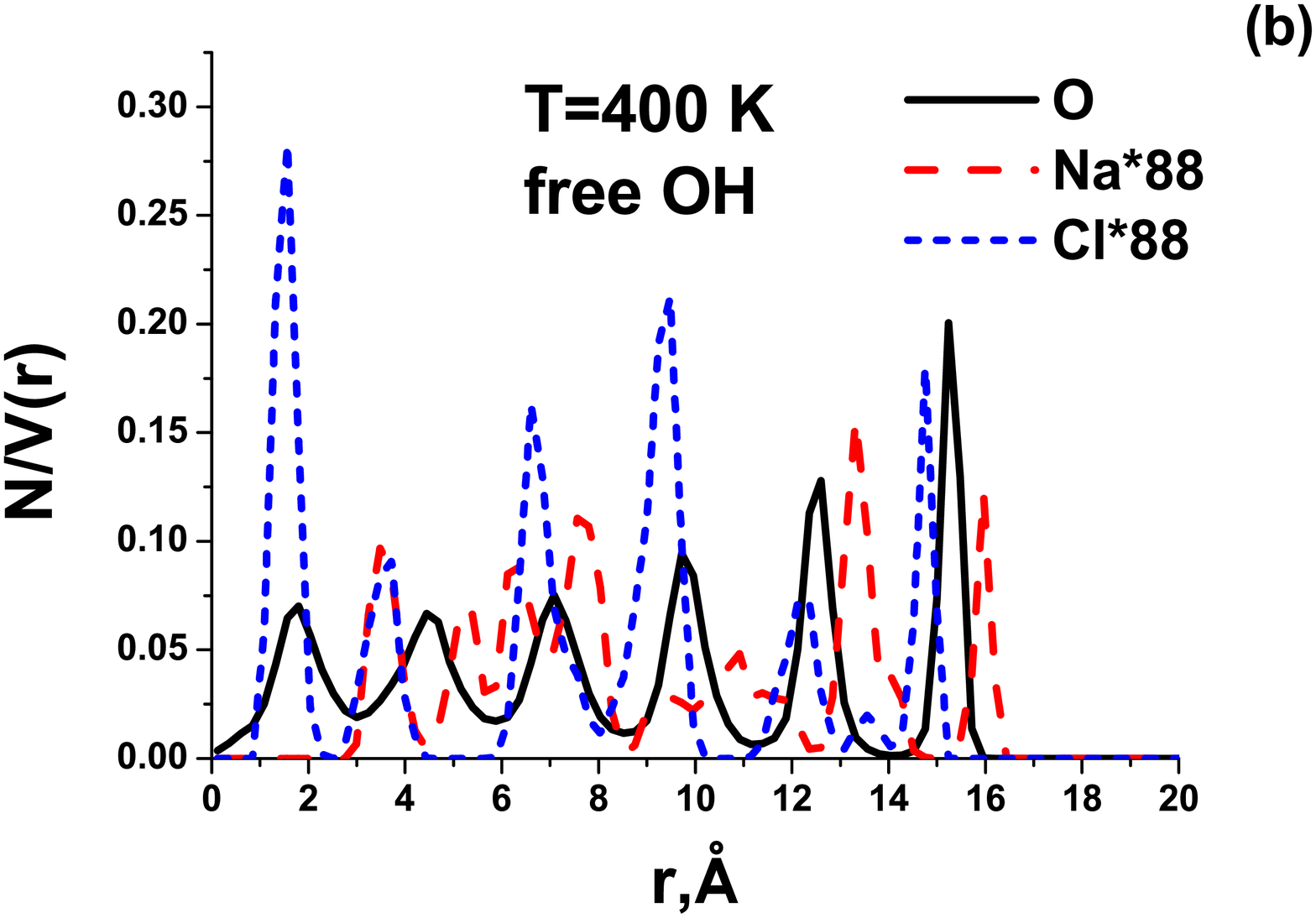}%

\caption{\label{fig:den-nacl-free-OH} Radial density distribution
of different species inside the asbestos fibers (a) with fixed
$OH$ groups and (b) with free $OH$ groups. $T=400 K$}
\end{figure}

In conclusion of this subsection we can see that the water
dynamics inside the fibers is very slow and therefore we need to
use very high temperatures (above $600 K$). However, asbestos tube
becomes unstable at such temperatures and we have to maintain it
rigid during the simulations. Although it affects the results only
small quantitative changes are detected. Therefore we believe that
our results give reasonable qualitative description of the
behavior of water inside the asbestos fibers.

\subsection{Sodium Chloride solution inside the asbestos fiber}

We proceed with study of sodium chloride solution inside the
asbestos fibers. $40$ randomly chosen molecules of water were
substituted by $Na^+$ and $Cl^-$ ion pairs and the behavior of the
solution was studied. An example snapshot of the system is given
in Fig. ~\ref{fig:water-nacl}.

\begin{figure}
\includegraphics[width=6cm, height=6cm]{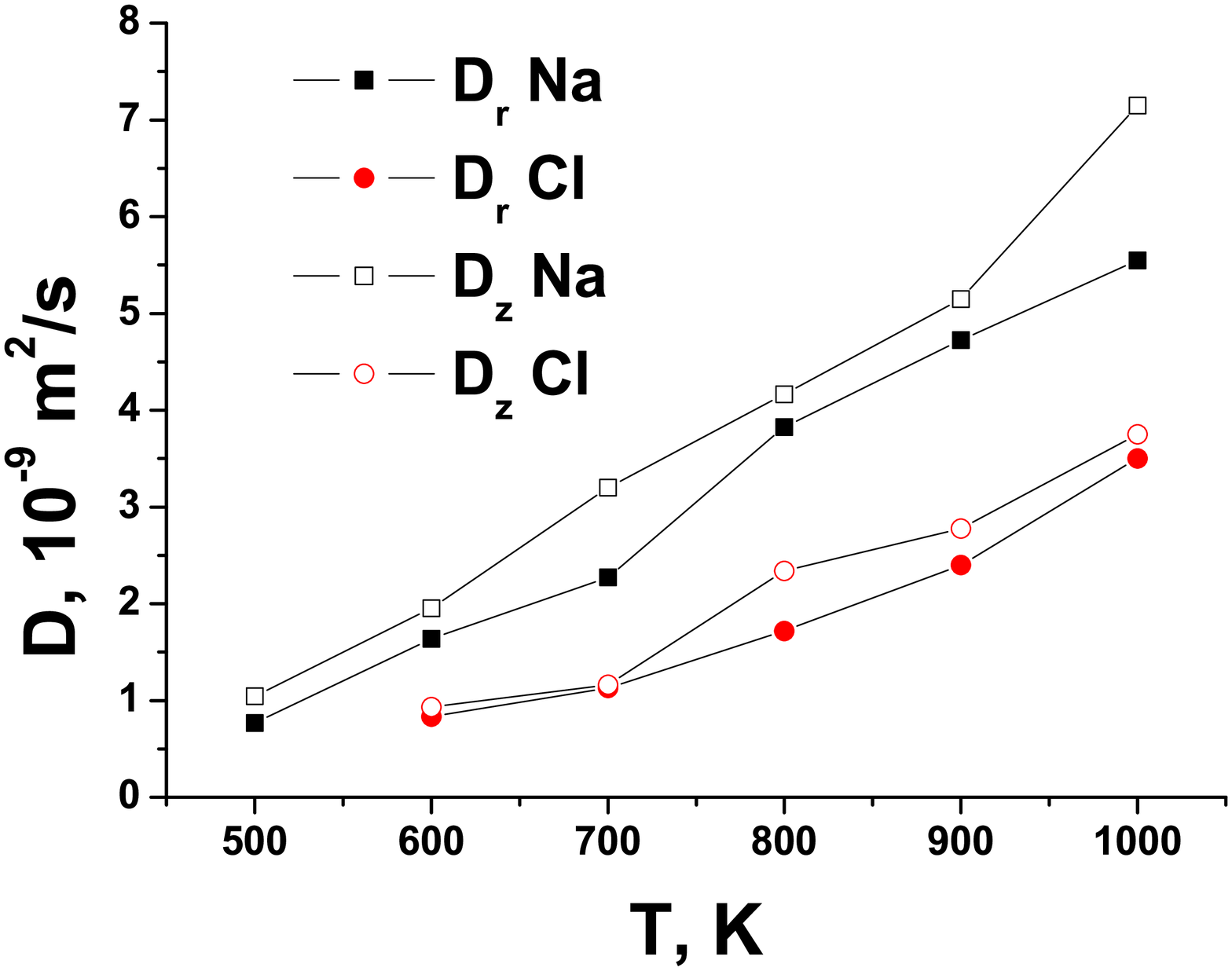}%

\caption{\label{fig:diff-nacl-OH} Diffusion coefficients of sodium
and chlorine ions inside the asbestos fiber. Radial and along the
tube axis coefficients are shown.}
\end{figure}

First, we study the radial density profiles of different species
inside the tube and we do not observe any influence on the water
structure inside the fibers due to sodium chloride introduction
which means that the system can be considered as a solution.

The number density profiles of different species of the solution
at different temperatures are shown in Fig. ~\ref{fig:den-nacl}.
One can see that the system demonstrates the layering of the
species with different charges. For example, if one considers the
case of $T=1000K$ the closest to the wall of the tube (large $r$)
is the layer containing oxygens and chlorine ions, i.e. negatively
charged species. The next layer is made of positively charged
sodium. After that another oxygen and chlorine layer is observed
and one more layer of sodium. At $r \leq 8 \AA$ sodium ions do not
show strong structuring while the densities of water and chlorine
are still modulated. A peak of sodium is observed at the tube
axis, however, this peak can be related to extremely small volume
$V=2 \pi r^2 dr H$ close to the tube axe and it can disappear in
the thermodynamic limit.

Qualitatively similar picture is observed for other temperatures.
As the temperature decreases the density modulations become more
pronounced and at $T=600 K$ all species demonstrate the density
modulations in the whole range of $r$.

We also compare the density profiles in the case of fixed and free
hydroxyl groups. The density profiles are shown in Fig.
~\ref{fig:den-nacl-free-OH}. One can see that like in the case of
pure water, making $OH$ groups free makes the system more
structured, but no strong qualitative changes appear. It allows us
to conclude that the system with rigid tube leads to qualitatively
correct behavior of the solution inside the fiber.

Fig. ~\ref{fig:diff-nacl-OH} shows the temperature dependence of
radial and axial diffusion coefficient of sodium and chlorine. One
can see that both sodium and chlorine have larger diffusion
coefficients then water. The diffusion coefficients of sodium
becomes zero at $T < 500 K$ and of chlorine at $T < 600K$.

\section{Conclusions}

We consider the behavior of water and sodium chloride solution
confined in lizardite asbestos nanotube which is a typical example
of hydrophilic confinement. Confining water into asbestos tube
strongly affects the dynamical properties. The diffusion
coefficient drops about two orders of magnitude comparing to the
bulk case. Like in the case of other hydrophilic pores water in
asbestos tubes experiences vitrification rather then
crystallization upon cooling. The density profiles are strongly
affected by the geometry of the tube. The modulations of the
density spread up to the whole tube region which is about $9 \AA$.

In the case of sodium chloride solutions we observe clear double
layer formation. It is shown that the closest to the wall of the
tube is the layer containing oxygens and chlorine ions, i.e.
negatively charged species while the next layer is made of
positively charged sodium. At temperatures below $600 K$ the
density profiles demonstrate very sharp peaks, i.e. the solution
becomes very structured in radial direction. It is interesting to
note that both sodium and chlorine have larger diffusion
coefficients then water.

\bigskip

\begin{acknowledgments}
Yu. F. thanks the Russian Scientific Center at Kurchatov Institute
and Joint Supercomputing Center of Russian Academy of Science for
computational facilities. The work was supported by Russian
Science Foundation (Grant No 14-12-00820).
\end{acknowledgments}


\end{document}